\documentclass[submission,copyright,creativecommons]{eptcs}
\providecommand{\event}{MeTRiD 2018} 
\usepackage{underscore}           

\usepackage{times}
\usepackage{amssymb,latexsym,amsfonts,amsmath}
\usepackage{graphicx}
\usepackage{subfig}

\newcommand{\N}{{\mathbb{N}}}
\newcommand{\R}{{\mathbb{R}}}
\newcommand{\tz}[1]{\times10^{#1}}
\newcommand{\MATLAB}{{\tt MATLAB}}
\newcommand{\OMNETPP}{{\tt OMNeT++}}
\newcommand{\SENSE}{{\tt SENSE}}
\newcommand{\SCOTS}{{\tt SCOTS}}
\newcommand{\CPP}{C++}
\newcommand{\CUDD}{{\tt CUDD}}

\newcommand{\exclude}[1]{}

\def\papertitle{SENSE: Abstraction-Based Synthesis of Networked Control Systems}

\title{\papertitle
	\thanks{This work was supported in part by the German Research Foundation (DFG) through the grant ZA 873/1-1.}
}

\author{Mahmoud Khaled
	\institute{
		Technical University of Munich\\
		Munich, Germany}
	\email{khaled.mahmoud@tum.de}
	\and
	Matthias Rungger
	\institute{
	Technical University of Munich\\
	Munich, Germany}
	\email{matthias.rungger@tum.de}
	\and
	Majid Zamani
	\institute{
		Technical University of Munich\\
		Munich, Germany}
	\email{zamani@tum.de}	
}

\def\authorrunning{
	M. Khaled, M. Rungger \& M. Zamani
}
\def\titlerunning{\papertitle}

\begin{document}
\maketitle

\begin{abstract}
While many studies and tools target the basic stabilizability problem of networked control systems (NCS), nowadays modern systems require more sophisticated objectives such as those expressed as formulae in linear temporal logic or as automata on infinite strings.
One general technique to achieve this is based on so-called symbolic models, where complex systems are approximated by finite abstractions, and then, correct-by-construction controllers are automatically synthesized for them.
We present tool \SENSE{} for the construction of finite abstractions for NCS and the automated synthesis of controllers. Constructed controllers enforce complex specifications over plants in NCS by taking into account several non-idealities of the communication channels.

Given a symbolic model of the plant and network parameters, \SENSE{} can efficiently construct a symbolic model of the NCS, by employing operations on binary decision diagrams (BDDs).
Then, it synthesizes symbolic controllers satisfying a class of specifications.
It has interfaces for the simulation and the visualization of the resulting closed-loop systems using \OMNETPP{} and \MATLAB.
Additionally, \SENSE{} can generate ready-to-implement VHDL/Verilog or C/C++ codes from the synthesized controllers.
\end{abstract}

\section{Introduction}
Networked control systems (NCS) combine physical components, computing devices, and communication networks all in one system forming a complex and heterogeneous class of so-called cyber-physical systems (CPS). 
NCS have attracted significant attention in the past decade due to their flexibility of deployment and maintenance (especially when using wireless communications).
However, numerous technical challenges are present due to the wide range of uncertainties within NCS, introduced by their unreliable communication channels. 
This includes time-varying communication delays,  packet dropouts,  time-varying sampling/transmission intervals, and communication constraints (e.g. scheduling protocols). 

Many studies deal with subsets of the previously mentioned imperfections targeting the basic stabilizability problem of NCS \cite{HeemelsStabilNCSchapter,CloostermanetalStabilNCStimevar,WouwetalDiscStabilNCS,DNesicDLiberzonUniFrameworkNCS}.  
However, nowadays CPS require more sophisticated specifications, including objectives and constraints given, for example, by formulae in linear temporal logic (LTL) or omega-regular languages \cite{CBaierPrincipModelChecking}.

A well-known approach to synthesize controllers enforcing such complex specifications is based on so-called symbolic models  (a.k.a. discrete abstractions) \cite{PTabuadaVCHSSymbolic,MZ,MZamanietalSCNonlinNoStabilAssump,GReissigetalFRRTAC}.
A given plant (i.e. a physical process described by a set of differential equations) is approximated by a symbolic model, i.e., a system with finite state and input sets. 
As the model is finite, algorithmic techniques from computer science \cite{WThomasSynthesisStrategiesinfGames,MalerPnueliSifakis95} are applicable to automatically synthesize discrete controllers enforcing complex logic specifications over symbolic models. 
Finally, those discrete controllers (a.k.a. symbolic controllers) can be refined to hybrid ones enforcing the same properties over original concrete systems.

In this paper, we present \SENSE{}, a tool for the automated synthesis of controllers for NCS.
It automatizes the construction of symbolic models of NCS, given symbolic models of plants in them and the network parameters, and then synthesizes controllers enforcing a class of temporal logic specifications over them.
Additionally, it generates automatically C/\CPP{} or VHDL/Verilog codes that are ready for implementation and, thus, providing an end-to-end automated synthesis approach for NCS.

\section{Symbolic models and controller synthesis for NCS}
\label{LBLSECSYMCONT}

\begin{figure}
	\centering
	\includegraphics[width=0.70\textwidth]{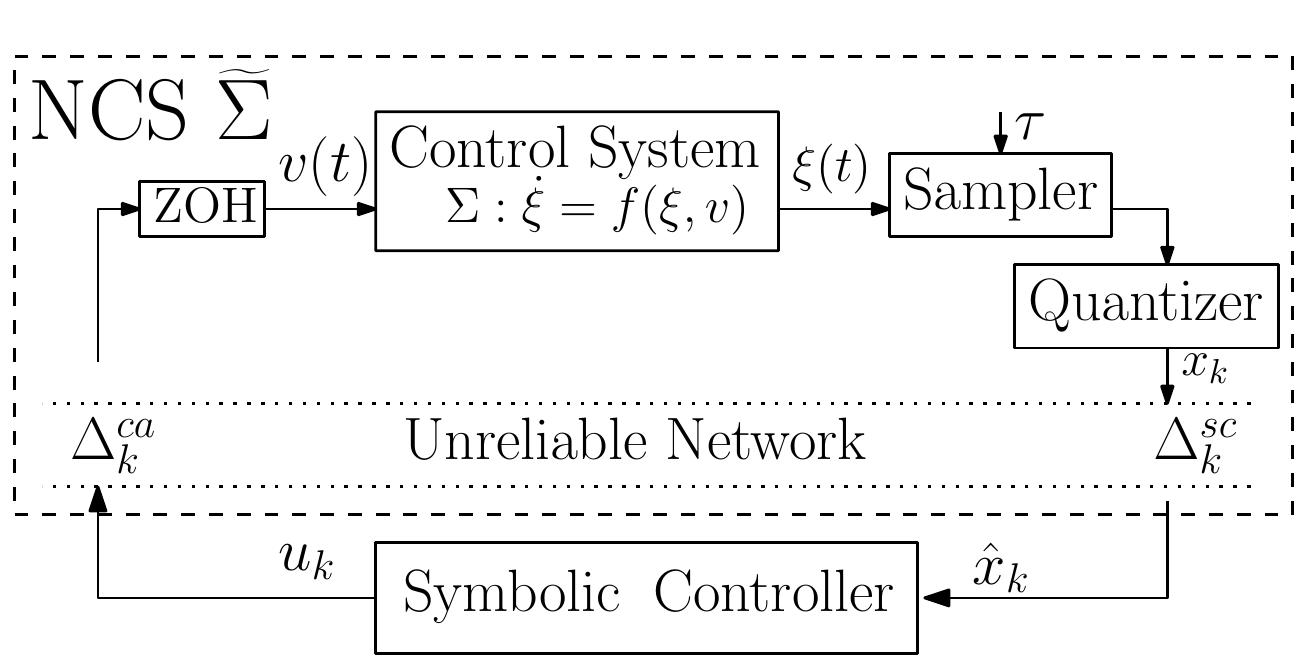}
	\caption{Structure of the NCS and the symbolic controller.}
	\label{FIGNCSDIAGRAM}
\end{figure}

Figure \ref{FIGNCSDIAGRAM} depicts the structure of NCS supported in the tool \SENSE{}, which follows the general NCS structure reported in \cite{WouwetalDiscStabilNCS} and \cite{zamani}.
The NCS $\widetilde{\Sigma}$ consists of the following components: 
(1) a control system, often referred to as plant, representing a physical process to be controlled; 
(2) two non-ideal communication channels transferring state information and control inputs from/to the plant; and 
(3) a remote digital (symbolic) controller enforcing some complex specifications
on the plant by taking into account the imperfections of the communication channels. 

The general (possibly nonlinear) control systems 
$\Sigma$ 
is usually modeled by a differential equation of the form $\dot{\xi}(t) = f(\xi(t), \upsilon(t))$ whose state $\xi(t) \in \R^n$ is measured, every sampling period $\tau$, by a sensor/sampler, that is followed by a quantizer.
The input signal $\upsilon$ belongs to the set $\mathcal{U}$, which is a subset of the set of all functions of time from $]a,b[ \subseteq \R$ to $\R^m$.
We denote by $\xi_{x\upsilon}(t)$ the state reached at time $t$ under the input $\upsilon$ and started from the initial condition $x=\xi_{x\upsilon}(0)$. We denote by
$x_k$ the sampled and quantized state, and $k \in \N_0$ is an index referring to the sampling time $k\tau$.
Additionally, a zero-order-hold (ZOH) helps sustaining an inter-sampling continuous-time constant control input $\upsilon(t)$.

A communication network connects the plant to a symbolic controller and introduces the time-varying sensor-to-controller and controller-to-actuator delays $\Delta_k^{sc}$ and $\Delta_k^{ca}$, respectively. 
A formal definition of symbolic controllers is given in \cite{GReissigetalFRRTAC}. 
We consider the delays to be bounded and to take the form of integer multiples of the sampling time $\tau$, meaning $\Delta_k^{sc} := N_k^{sc}\tau$ and $\Delta_k^{ca} := N_k^{ca}\tau$, where $N_k^{sc} \in [N_{\min}^{sc};N_{\max}^{sc}]\subset \N$, $N_k^{ca} \in [N_{\min}^{ca};N_{\max}^{ca}]\subset\N$, and $N_{\min}^{sc}, N_{\max}^{sc}, N_{\min}^{ca}, N_{\max}^{ca} \in \N$.

Having a quantizer before the network as in Figure \ref{FIGNCSDIAGRAM} and considering time-varying communication delays with upper and lower bounds, as showed in \cite[and references therein]{mkhaledallerton16,MZamanietalSCFiniteAbsNCS,zamani}, one can readily consider four types of network non-idealities: (i) quantization errors; (ii)  limited bandwidth; (iii) time-varying communication delays; and (iv) packet dropouts as long as the maximum number of consecutive dropouts over the network is bounded \cite{HeemelsStabilNCSchapter}.

\subsection{Symbolic Control of NCS}
We use the symbolic approach to design controllers enforcing given logic specifications over the plants in NCS while taking the network imperfections into account. Here, we use a notion of system, which is introduced in \cite{PTabuadaVCHSSymbolic}, as a unified modeling framework for both continuous systems as well as their finite abstractions.
A system is a tuple $S=(X, X_0, U, F)$ that consists of
a state set $X$;
a set of initial states $X_0 \subseteq X$;
an input set $U$; and 
a transition relation $F  \subseteq X \times U \times X$.
We also use $F$ as a map to denote the set of post-states of $x$ which is $F(x,u) = \{ x' \in X \vert (x,u,x') \in F \}$.

Let $S_\tau(\Sigma) = (\R^n, \R^n, U_\tau, F_\tau)$ denote the system that captures the evolution of $\Sigma$ at each sampling time $\tau$. Set
$U_\tau = \{\upsilon: [0,\tau[ \rightarrow \R^m \mid \upsilon(0) \in \R^m \text{, and } \upsilon(t) = \upsilon(0), \text{ for all } 0 \leq t < \tau \}$ is a an input set of constant curves over intervals of length $\tau$.
A transition $(x_\tau, \upsilon_\tau, x'_\tau)$ belongs to $F_\tau$ if and only if there exists a trajectory $\xi_{x_\tau \upsilon_\tau} : [0, \tau] \rightarrow \R^n$ in $\Sigma$ such that $\xi_{x_\tau \upsilon_\tau}(\tau) = x'_\tau$.
 
Let $S_q(\Sigma) = (X_q, X_{q,0}, U_q, F_q)$, where $X_q$ is a finite cover of $X_\tau$ and $U_q$ is a finite subset of $U_\tau$, denotes the symbolic model of $S_\tau(\Sigma)$ if there exists a \emph{feedback refinement relation} (FRR) $Q \subseteq X_\tau \times X_q$ from $S_\tau(\Sigma)$ to $S_q(\Sigma)$.
Interested readers can find more details about FRR and the construction of $F_q$ in \cite{GReissigetalFRRTAC}. System
$S_q(\Sigma)$ can then be used to synthesize controllers to enforce given logic specifications over $\Sigma$.

For NCS, one can follow the same methodology.
A system $\widetilde{S}_\tau(\Sigma)$, representing $S_\tau(\Sigma)$ in the network environment with the network imperfections, is derived from $S_\tau(\Sigma)$.
Then, $\widetilde{S}_q(\Sigma)$, which represents a symbolic model of the overall NCS, is constructed from  $\widetilde{S}_\tau(\Sigma)$ such that there exists a FRR $\widetilde{Q}$ from $\widetilde{S}_\tau(\Sigma)$ to $\widetilde{S}_q(\Sigma)$.
The process of constructing $\widetilde{S}_q(\Sigma)$ directly from $\widetilde{S}_\tau(\Sigma)$ is rather very complex since $\widetilde{S}_\tau(\Sigma)$ is a high-dimensional representation of $S_\tau(\Sigma)$ that includes network imperfections.
Instead, \SENSE{} employs the results in \cite[Theorem 4.1]{mkhaledallerton16} to construct $\widetilde{S}_q(\Sigma)$ directly from $S_q(\Sigma)$ while preserving some FRR, without going through the construction of $\widetilde{S}_\tau(\Sigma)$.
This construction is achieved by leveraging an operator $\mathcal{L}$ to ${S}_q(\Sigma)$ as introduced in \cite{mkhaledallerton16,MZ}.
More specifically, $\widetilde{S}_q(\Sigma) := (\widetilde{X}_q, \widetilde{X}_{q, 0}, U_q, \widetilde{F}_q)$ is derived as $ \widetilde{S}_q(\Sigma) = \mathcal{L}(S_q(\Sigma), N^{sc}_{\min}, N^{sc}_{\max}, N^{ca}_{\min}, N^{ca}_{\max})$, where
\begin{itemize}	

	\item $\widetilde{X}_q = \{X_q \cup q\}^{N^{sc}_{\max}} \times U_{q}^{N^{ca}_{\max}} \times [N^{sc}_{\min}; N^{sc}_{\max}]^{N^{sc}_{\max}} \times [N^{ca}_{\min}; N^{ca}_{\max}]^{N^{ca}_{\max}}$, where $q$ is a dummy symbol representing the lack of state due to channel initialization;

	\item $\widetilde{X}_{q,0} = \{(x_0,q,\ldots,q,u_0,\ldots, u_0,N^{sc}_{\max},\ldots,N^{sc}_{\max},N^{ca}_{\max},\ldots,N^{ca}_{\max})$  s.t. $x_0 \in X_{q,0}$ and $u_0 \in U_{\tau} \}$;

	\item A transition $((x_1,\ldots,x_{N^{sc}_{\max}},   u_1,\ldots,u_{N^{ca}_{\max}},  \widetilde{N}_1,\ldots,\widetilde{N}_{N^{sc}_{\max}}, \widehat{N}_1,\ldots,\widehat{N}_{N^{ca}_{\max}}),  u,\\ (x',x_1,\ldots,x_{N^{sc}_{\max}-1},   u,u_1,\ldots,u_{N^{ca}_{\max}-1},  \widetilde{N},\widetilde{N}_1,\ldots,\widetilde{N}_{N^{sc}_{\max}-1}, \widehat{N},\widehat{N}_1,\ldots,\widehat{N}_{N^{ca}_{\max}-1}))$ belongs to $\widetilde{F}_q$, for all $\widetilde{N} \in [N^{sc}_{\min}; N^{sc}_{\max}]$ and for all $\widehat{N} \in [N^{ca}_{\max}; N^{ca}_{\max}]$, if there exists a transition $x_0 \overset{u_{N^{ca}_{\max} - j*}}{\underset{\tau}{\longrightarrow}} x'$ in $F_q$, where $j*$ is a time-shifting index defined in \cite{mkhaledallerton16}.	
\end{itemize}

Symbolic controllers are generally finite systems that accept states $\hat{x}_k$ as inputs and produce $u_k$ as outputs to enforce a given specification over the plants in NCS.
The tool \SENSE{} natively supports safety, reachability, persistence, and recurrence specifications given as the LTL formulae $\square \varphi_S$, $\Diamond \varphi_T$, $\Diamond\square \varphi_S$, $\square\Diamond \varphi_T$, respectively, where $\varphi_T$ and $\varphi_S$ are the predicates defining, respectively, some target and safe sets $T$ and $S$.
In the next section, we introduce three main modules inside \SENSE:
\begin{itemize}
	\item[(1)] The symbolic model construction engine: it is responsible for constructing the symbolic model of the NCS from a symbolic model of the plant in it;
	\item[(2)] The fixed-point operations engine: it is responsible for synthesizing correct-by-construction symbolic controllers enforcing the aforementioned LTL properties over the NCS;
	\item[(3)] Various tools to analyze and simulate the closed-loop NCS, and to automatically generate code of the synthesized control software.
\end{itemize}

\section{Structure of the tool \SENSE{}}
\label{LBLSECSENSETOOL}
The tool \SENSE{} is an open-source \emph{extensible} framework that provides a base for further research on the modeling, abstraction, and controller synthesis of NCS.
The research on the symbolic control for NCS is still ongoing \cite{zamani,mkhaledallerton16} and the tool is developed to help researchers interested in automated synthesis of NCS.
 
For the construction of symbolic models of NCS, \SENSE{} has a software engine that supervises the construction procedure.
The engine follows certain customizable construction rules that describe how the symbolic model of the NCS is constructed from the symbolic model of the plant. 
The construction rules depend on the sensor-to-controller and controller-to-actuator delays $\Delta_k^{sc}$ and $\Delta_k^{ca}$, respectively. 
Construction rules (implemented in terms of \CPP{} classes) can be readily provided for any class of NCS.
However, due to the imposed restriction in the implementation phase of the synthesized controllers, following the existing theory in ~\cite{mkhaledallerton16,zamani}, tool \SENSE{} is currently available with only the construction rules for a class of NCS called prolonged-delay NCS (introduced later). 
Nevertheless, due to the modular design of \SENSE{}, it is straightforward to define extra construction rules, e.g. taking time-varying delays into account, once the theory behind them is available.

The tool \SENSE{} provides ready-to-use fixed-point routines that operate on the constructed symbolic models to synthesize controllers enforcing any of those specifications introduced in Section \ref{LBLSECSYMCONT}.
It is also equipped with two interfaces to access the synthesized controllers. 
The first one is via \MATLAB{} and allows accessing the synthesized controller as well as performing closed-loop simulations, in which, the plant is being remotely controlled by the synthesized controller over the network. 
We include a second interface using \OMNETPP{} \cite{OMNETPP} that provides a powerful visualization-framework and realistic network models.
The interface of \OMNETPP{} allows accessing the synthesized controller to perform realistic network simulations as well as visualizing the closed-loop behavior.

Along with the core engines in \SENSE, a rich set of helper tools is developed and made available for users to analyze and simulate the resulting symbolic models and the synthesized symbolic controllers.
One of those tools allows for code generation of the synthesized controllers as C/\CPP{} or VHDL/Verilog codes to cover both software and hardware implementations and provide an end-to-end solution. 

\subsection{The symbolic model construction engine}
\label{LBLSECENGINE}

\begin{figure}
	\centering
		\includegraphics[width=0.80\textwidth]{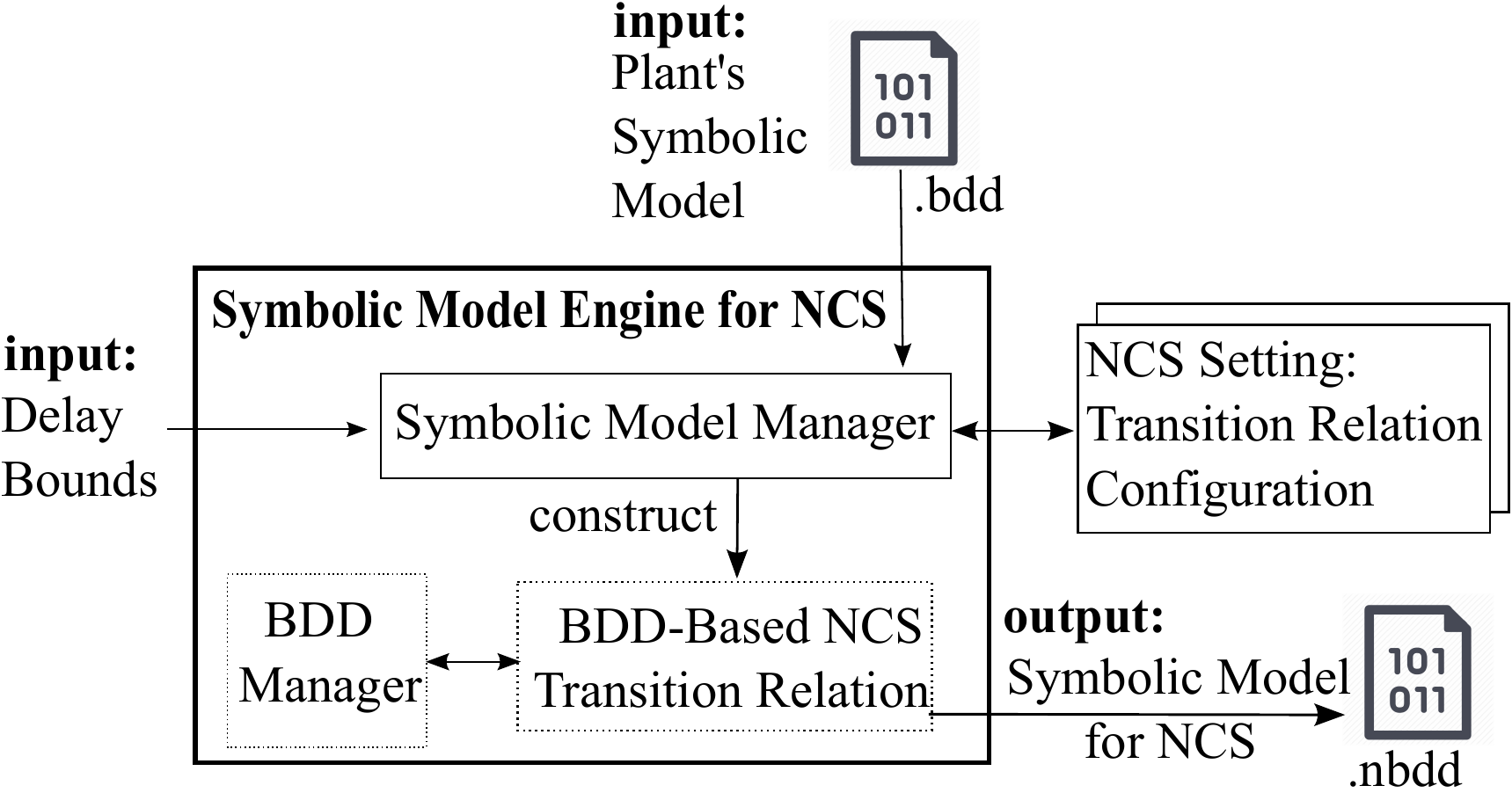}
	\caption{Structure of the engine for constructing symbolic models of NCS.}
	\label{FIGSENSEENGINE}
\end{figure}

A straightforward implementation of $\mathcal{L}$-operator can be achieved using element-by-element expansion of the state set of $S_q(\Sigma)$.
However, this is computationally inefficient and consumes a large amount of memory.

Symbolic models, which can be considered as labeled transition systems, are easily represented by boolean functions and encoded inside BDDs. Tool \SENSE{} expands the BDD representing $S_q(\Sigma)$ to construct a BDD representing $\widetilde{S}_q(\Sigma)$.
Symbolic operations on BDDs play the major role in order to construct symbolic models of NCS from those of their plants.
Using the \CUDD{} library \cite{CUDDLIBMANUAL}, \SENSE{} employs such operations to efficiently handle computation/memory complexities when implementing the $\mathcal{L}$-operator.

Figure \ref{FIGSENSEENGINE} depicts the structure of the engine for constructing symbolic models of NCS.
The engine takes the symbolic model of the plant as its input. 
Users can construct such models using the existing tool \SCOTS{} \cite{SCOTS}, which is provided as a library inside the tool \SENSE.
The constructed model is then given to the engine in the form of a BDD file (i.e., a file storing the BDD).
The delay bounds within the NCS should also be provided.
The engine then starts expanding the symbolic model of the plant.
Specifically, the transition relation of $\widetilde{S}_q(\Sigma)$ is crafted by Cartesian-product-like expansion of the transition relation of $S_q(\Sigma)$.
This is done with the help of a BDD-Manager provided by the CUDD library as follows:
\begin{itemize}
	\item[(1)] adding extra binary variables to accommodate for the traveling state and input packets through the communication channels, and their delays; 	
	\item[(2)] to construct $\widetilde{X}_q$, \SENSE{} expands $X_q$ with the dummy symbol $q$ and then performs Cartesian products with $U_q$ and the set of delays in both channels;
	\item[(3)] to construct $\widetilde{F}_q$, \SENSE{} performs BDD operations on $F_q$ using the provided construction rules described as \CPP{} classes.
\end{itemize}


\subsubsection{The prolonged-delay NCS}
Note that for constructing symbolic models of any NCS (using the operator $\mathcal{L}$) and synthesizing symbolic controllers enforcing any of the properties introduced in Section \ref{LBLSECSYMCONT} over them, we only require the delays in both channels of the network to be integer multiples of the sampling period $\tau$ with some lower and upper bounds; see \cite{mkhaledallerton16,zamani} for more details. 

The tool \SENSE{} can handle the construction of symbolic models and the synthesis of their controllers for any class of NCS.
However, in order to refine the controllers and apply them to the concrete NCS, the current theory proposed in \cite{zamani} requires that the upper and lower bounds of the delays to be equal at each channel. 
This implies that, in each channel, all packets are delayed by the same amount of time. 
This can be practically achieved by performing extra prolongation (if needed) of the delays suffered by the packets.  
For the sensor-to-controller branch of the network, this can be readily done inside the controller. 
The controller needs to have a buffer to hold arriving packets and keep them in the buffer until their delays reach the maximum. 
For the controller-to-actuator channel, the same needs to be implemented inside the ZOH. Interestingly, this prolongation results in less conservativeness in terms of the existence of symbolic controllers; see \cite[Lemma 6.1]{zamani} for more details.

Additionally, to refine the synthesized symbolic controllers, it is also required that $S_q(\Sigma)$ is deterministic. 
This can be fulfilled for a wide class of physical systems \cite{PTabuadaVCHSSymbolic} having some stability properties, and it does not require $\widetilde{S}_q(\Sigma)$ to be deterministic.
Nevertheless, the support of controller refinement for non-deterministic symbolic models of the plants is currently under development and will be incorporated in \SENSE{}. 


\subsection{Fixed-point computation for controller synthesis}
\label{LBLSECFP}
The tool \SENSE{} implements fixed-point algorithms for synthesizing controllers that enforce those specifications introduced in Section \ref{LBLSECSYMCONT}.
	Tool \SENSE{} uses a customized version of the controller synthesis engine provided in \SCOTS{} \cite{SCOTS}.
	In order to synthesize controllers that enforce any of the four specifications, the engine implements fixed-point algorithms that employ a {\tt Pre}-operation over state-input pairs of $\widetilde{S}_q(\Sigma)$.
	More specifically, given a symbolic model of the NCS $\widetilde{S}_q(\Sigma)$ and a set $Z \subseteq \widetilde{X}_q \times U_q$, the {\tt Pre}-operation of $\widetilde{X}_q$ is defined by  $pre(Z) = \{ (x,u) \in \widetilde{X}_q \times U_q  \vert \emptyset \neq \widetilde{F}_q(x, u) \subseteq \pi_{\widetilde{X}_q}(Z) \}$, where $\pi_{\widetilde{X}_q}$ is a projection map defined as $\pi_{\widetilde{X}_q}(Z) = \{ x \in \widetilde{X}_q \vert \exists_{u \in U_q} \; (x,u) \in Z \}$.
	Interested readers can find more details in \cite{SCOTS,PTabuadaVCHSSymbolic,GReissigetalFRRTAC}.
All steps in the algorithms, as well as the {\tt Pre}-operation, are implemented by BDD operations.
Users can employ the four main algorithms to describe different LTL specifications and to synthesize (possibly dynamic) symbolic controllers.
We demonstrate this, with examples, for two different LTL specifications later in Section \ref{LBLSECEXRESULTS}.

The target set $T$ and safe set $S$ are to be given as atomic propositions over the state space of the plant.
Such sets can be generated using \SCOTS{} and they are provided as BDD files.
\SENSE{} takes care of expanding those sets to be compatible with the symbolic model of the NCS.
Then, it synthesizes controllers to enforce the specifications over the original plants in NCS.
Controllers are, by construction, BDD objects and they are saved as BDD files.

\subsection{Simulation, analysis and code generation}
\label{LBLSECSIM}
The tool \SENSE{} provides two different interfaces to access the controllers, using \MATLAB{} and \OMNETPP.
For both interfaces, it provides a common \CPP{} layer to read the BDD-based synthesized controllers.

For \MATLAB, an \texttt{m}-script uses the \CPP{} layer to access the controllers while providing a set of \texttt{m}-script functions and classes to interface the synthesized controllers.
For almost all examples, \texttt{m}-scripts are provided to simulate the closed-loop behavior of the NCS. 
This includes simulating the plant's differential equation, and the network evolution.

In the interface for \OMNETPP, communication channels are modeled as random-delay channels for realistic simulations.
Visualizations up to 3 dimensions are supported in the visualization engine of \OMNETPP.

\SENSE{} comes equipped with the following tools that help analyzing the symbolic models as well as the synthesized controllers:
\begin{itemize}
	\item {\tt bdd2implement}: a tool to automatically generate C/C++ or VHDL/Verilog codes from the synthesized BDD-based controllers;
	\item {\tt bdd2fsm}: a tool to generate files following the FSM data format \cite{FSMFormat} or the comma-separated format of the transition relation of symbolic models. Such representation is used by many software packages to visualize graphs (e.g., the tool StateVis\cite{StateVis}). Users can use the tool to analyze or understand, visually, how the symbolic model behaves;
	\item {\tt bddDump}: a tool to extract the meta-data information stored inside the generated BDD files. This helps inspecting the original information about the plant such as sampling time, quantization parameters, or the binary variables;
	\item {\tt contCoverage}: a tool to provide fast terminal-based ASCII-art visualization of the coverage of the synthesized controllers. It is only possible up to 2 dimensions. To visualize the coverage of controllers with 3 dimensional inputs, users can still use \MATLAB.
	\item {\tt sysExplorer}: a tool to help testing the expanded transition relation or the synthesized controllers. It handles the BDD as IN-OUT box. Users can request information about specific transitions in the transition relation by providing the initial state and a sequence of inputs, and the tool responds with the post-states. The tool can be used to test the controller by providing some states to check their corresponding generated control inputs.
\end{itemize}

\subsubsection{Automated code generation of synthesized controllers}
\label{LBLBDDIMPLTOOL}
In \SENSE, the helper tool {\tt bdd2implement} is provided to automatically generate codes, targeting hardware and software implementations, of the synthesized controllers.
It starts by determinizing the symbolic controller, in case it is not deterministic, by simply selecting the first available input, for each state of the system.
Then, it converts the multi-output boolean function, representing the controller and encoded as a BDD, to many single-output functions.
Each single-output boolean function represents one bit of the binary-encoded output of the controller.
Then,
\begin{itemize}
	\item for C/C++ code generation, the boolean functions are dumped as C++ functions with {\tt bool} return values along with another function that collects and constructs the output of the controller.
	The C/C++ target-specific compiler takes care of converting such boolean functions to machine codes, implying that the controller is encoded as instructions of the target microprocessor. This, however, might violate the real-time requirements for some applications in which the time required to compute the control output, by executing the instructions, exceeds the real-time deadline. A more time-efficient (but not memory-efficient) implementation is to store the controller as a lookup-table in the memory of the target hardware and access it directly. The latter is straightforward but it is not currently added to the tool since most of the symbolic controllers of NCS have large sizes.
	
	\item for VHDL/Verilog code generation, the boolean functions are directly written as register transfer level (RTL) codes. An electronic design automation (EDA) tool takes care of converting the RTL code to a technology-mapped netlist. Then, a vendor-specific software finds the suitable mapping of logic-gates inside the target chip, in an operation called place-and-route.
\end{itemize}

\subsection{The work flow inside \SENSE}

\begin{figure}
	\centering
	\includegraphics[scale=0.65]{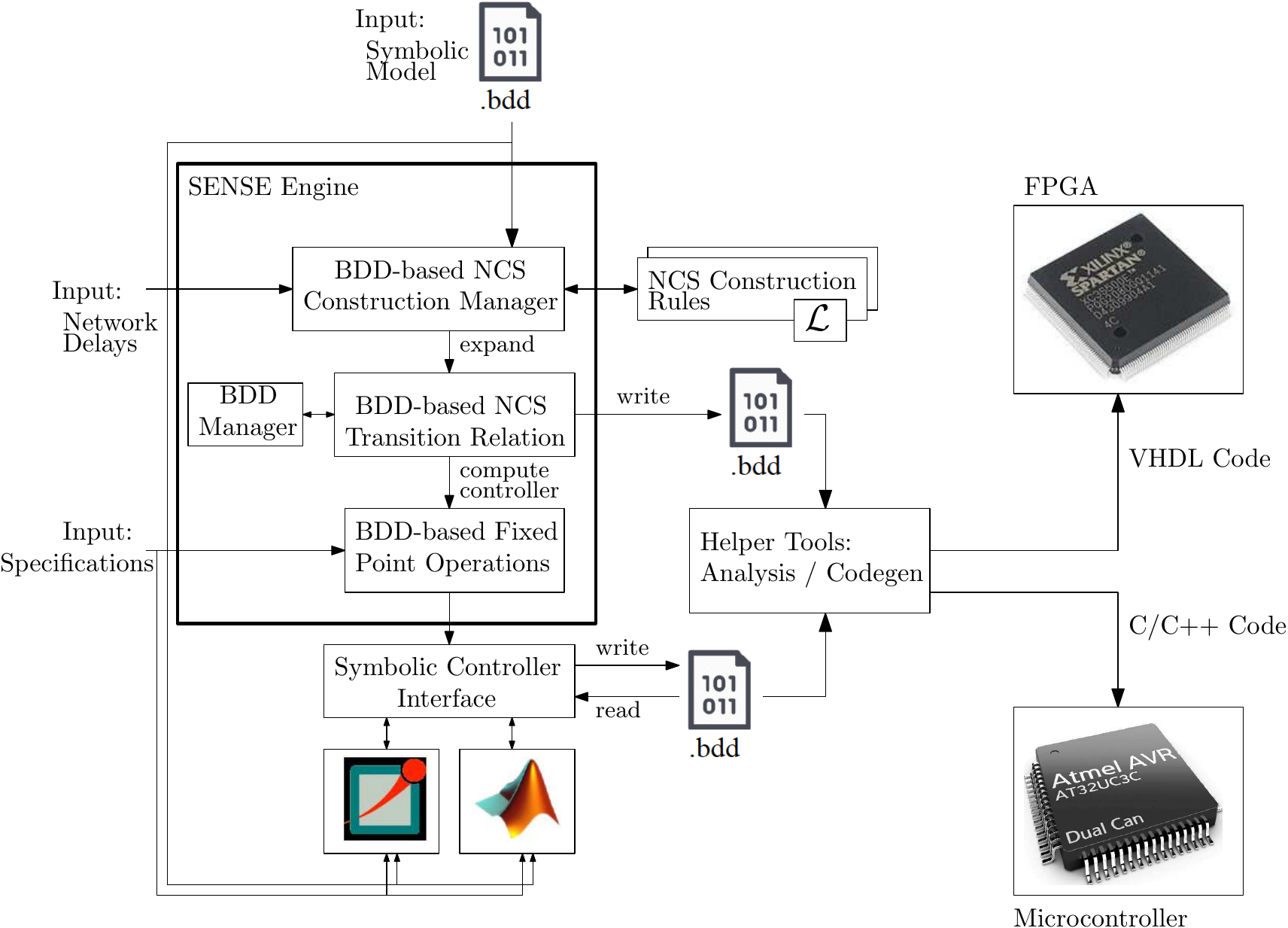}
	\caption{Work flow inside \SENSE: from specifications to ready-to-implement source codes.}
	\label{FIGSENSEWORKFLOW}
\end{figure}

Figure \ref{FIGSENSEWORKFLOW} depicts the complete structure and the work-flow of \SENSE.
As an input, the engine takes the symbolic model of the plant, the delay bounds within the NCS, and the specification.
Then, BDD files of the symbolic model of the NCS, following the construction rules, and its controller are generated.
The files are used by the helper tools for the analysis and code generation purposes.
The steps taken by the user to use \SENSE{} are summarized as follows:  
\begin{enumerate}
	\item The user computes a symbolic model of the plant as a BDD file using the included library \SCOTS;
	\item The user can extract information of the BDD file using the helper tool {\tt bddDump};
	\item The user can also visualize the transition relation of the symbolic model via the tool {\tt bdd2fsm} and a graph visualizing tool (e.g. StateVis);
	\item The user passes the BDD file representing the plant's symbolic model to \SENSE{}, along with the network delays and the specification;
	\item The tool \SENSE{} generates BDD files for the symbolic model of the NCS and the controller;
	\item The user can benefit from the tools {\tt sysExplorer} and {\tt contCoverage} to test the symbolic model of the NCS and its synthesized controller;
	\item The user can use the provided interfaces in \SENSE{}, in	combination with \MATLAB{} and/or \OMNETPP, to simulate and visualize the closed-loop behavior of the NCS;
	\item The user can use the tool {\tt bdd2implement} to generate the final controller for implementation. 
\end{enumerate}

\section{\SENSE{} in action: results and examples}
\label{LBLSECEXRESULTS}

We present several case studies where we construct symbolic models of NCS, namely, $\widetilde{S}_q(\Sigma)$ from the symbolic models of their plants, namely, $S_q(\Sigma)$ . 
We also provide two examples where dynamic controllers are synthesized.
The closed-loop is simulated using the interfaces with \MATLAB{} and \OMNETPP{} in \SENSE.
Code generation is achieved using the tool {\tt bdd2implement} in \SENSE.
All examples are done in a PC (Intel Xeon-E5-1620 3.5GHz 256GB RAM). Users can find the download links of \SENSE, the manual, a quick-start guide at \url{www.hcs.ei.tum.de}.

\subsection{Construction of symbolic models of NCS}
\begin{table*}[t]	
	\small
	\centering
	\caption{Results for constructing symbolic models of prolonged-delay NCS using those of their plants.}
	\label{TBLMODELCONSDETAILS}
	\scalebox{0.65}{	
		\begin{tabular}{lll|llllllllll}
			\hline
			Case Study 		 		& $\vert S_q(\Sigma)\vert$ &								& (2,2) 	  & (2,3)  	    & (2,4) 	  & (2,5)		& (3,2) 	  & (3,3)  	    & (3,4) 	  & (3,5) 	    & (4,2) 	  & (4,3) \\ \hline
			
			\textsf{DI} 	 		& 2039    		   & $\vert \widetilde{S}_q(\Sigma)\vert$  	& 14096 	  & 56336 	    & 225296 		& 901136      & 22832 	    & 91184       & 364592     & 1.45$\tz{6}$ & 38340 & 152964 \\
			&         		   & Time (sec)	    						& $< 1$ 	  & $< 1$  	    & $< 1$ 	  & $< 1$		& $< 1$ 	  & $< 1$  	    & $< 1$ 	  & $< 1$ 	    & $< 1$ 	  & $< 1$  	    \\
			&         		   & Memory (KB)    						& 2.0         & 2.4         & 3.1         & 2.9         & 3.0         & 2.9         & 3.1         & 3.2         & 5.2         & 4.3         \\ \cline{3-13}
			
			\textsf{Robot} 	 		& 29280  		   & $\vert \widetilde{S}_q(\Sigma)\vert$  				  & 4.1$\tz{6}$ & 6.5$\tz{7}$ & 1.04$\tz{9}$& 1.67$\tz{10}$& 3.4$\tz{7}$ & 5.4$\tz{8}$ & 8.7$\tz{9}$& 1.4$\tz{11}$& 2.8$\tz{8}$ & 4.6$\tz{9}$\\
			&         		   & Time (sec)	    						& $< 1$ 	  & $< 1$  	    & $< 1$ 	  & $< 1$       & $< 1$       & $< 1$       & $< 1$       & $< 1$       & 1.4         & 1.6         \\
			&         		   & Memory (KB)    						& 15          & 14          & 17          & 16          & 16          & 21          & 22          & 19          & 35          & 33          \\ \cline{3-13}
			
			\textsf{Jet} 			& $9.0\times10^5$  & $\vert \widetilde{S}_q(\Sigma)\vert$  	& 1.5$\tz{11}$& 1.0$\tz{13}$& 6.5$\tz{14}$& 4.2$\tz{16}$& 7.2$\tz{12}$& 4.6$\tz{14}$& 2.9$\tz{16}$& 1.9$\tz{19}$& 3.3$\tz{14}$& 2.1$\tz{16}$ \\
			&         		   & Time (sec)	    						& 1970  	  & 1637  	    & 1674 	      & 2172		& 3408   	  & 1772  	    & 2111 	      & 7107 	    & 4011 	      & 2854  	    \\
			&         		   & Memory (KB)    						& 4323.3      & 2374.2      & 2389.1      & 2392.6      & 3683.2      & 4098.2      & 3317.2      & 3582.6      & 5894.3      & 4784.5      \\ \cline{3-13}
			
			\textsf{DC-DC} 			& $3.8\times10^6$  & $\vert \widetilde{S}_q(\Sigma)\vert$  	& 8.9$\tz{7}$ & 1.8$\tz{8}$ & 3.6$\tz{8}$ & 7.1$\tz{8}$ & 5.2$\tz{8}$ & 1.0$\tz{9}$ & 2.0$\tz{9}$ & 4.1$\tz{9}$ & 3.0$\tz{9}$ & 6.0$\tz{9}$ \\
			&         		   & Time (sec)	    						& 672 	      & 681  	    & 530   	  & 1131		& 13690 	  & 10114  	    & 9791   	  & 10084 	    & 139693 	  & 137648      \\
			&         		   & Memory (KB)    						& 3347.2      & 3145.0      & 3176.7      & 2784.8      & 10875.2     & 11543.8     & 11572.0     & 11592.1     & 34169.2     & 37852.6     \\ \cline{3-13}
			
			
			
			\textsf{Vehi}			& $1.9\times10^7$& $\vert \widetilde{S}_q(\Sigma)\vert$  		& 4.2$\tz{12}$& 2.7$\tz{14}$& 1.7$\tz{16}$& 1.1$\tz{18}$& 2.3$\tz{14}$& 1.4$\tz{16}$& 9.3$\tz{17}$& 5.9$\tz{19}$& 1.3$\tz{16}$& 7.94$\tz{17}$\\
			&         		   & Time (sec)	    						& 273.3 	  & 285  		& 238.4 	  & 173 		& 22344 	  & 54919  	    & 27667 	  & 36467 	    & 39065 	  & 145390  	\\
			&         		   & Memory (KB) 							& 1638.4      & 1945.6      & 1843.2      & 1945.6      & 23040       & 40652.8     & 30208       & 22425.6     & 21094.4     & 36556.3		\\ \cline{3-13}
			\textsf{Inver}  		& $4.3\times10^8$& $\vert \widetilde{S}_q(\Sigma)\vert$  		& 2.4$\tz{14}$& 1.5$\tz{16}$& 1.0$\tz{18}$& 6.5$\tz{19}$& 4.6$\tz{16}$& 2.9$\tz{18}$& 1.9$\tz{20}$& 1.2$\tz{22}$& 9.2$\tz{16}$& 5.8$\tz{20}$ \\
			&         		   & Time (sec)	    						& 361.4		  & 349.1		& 340.9	  	  & 347.8		& 942		  & 793			& 1218		  & 910 		& 58411	  	  & 57110        \\
			&         		   & Memory (KB) 							& 723.1		  & 808.4		& 349.6		  & 1012.7		& 2958		  & 2840		& 3104 		  & 2898		& 35907		  & 35628        \\
			
			\hline											 			 			
		\end{tabular}
	}
\end{table*}

First, symbolic models of the plants in NCS are constructed and stored as BDD files. 
The BDD files are fed as inputs to the engine of \SENSE{} along with NCS delay bounds to construct symbolic models of NCS. 
Table \ref{TBLMODELCONSDETAILS} summarizes the results for different case studies (left-side column) and different network delays (top row) given as pairs $(a,b)$, where the delay of the sensor-to-controller channel is upper bounded by $a\tau$ and the delay of the controller-to-actuator channel is upper bounded by $b\tau$.
For each case study, we show the size of the symbolic model of the plant (i.e. number of transitions and denoted by $\vert S_q(\Sigma) \vert$). 
Then, for each network delay, we show the size of the resulting symbolic model of NCS (denoted by $\vert \widetilde{S}_q(\Sigma) \vert$), the time in seconds required to construct it, and the memory in KB used to store it.

In Table \ref{TBLMODELCONSDETAILS}, we consider several different dynamics for the plants in the NCS: 
a double integrator (denoted by \textsf{DI}), 
a fully actuated robot (denoted by \textsf{Robot}), 
a jet engine (denoted by \textsf{Jet}), 
a Boost DC-DC Converter (denoted by \textsf{DC-DC}), 
a vehicle dynamic (denoted by \textsf{Vehi}), 
and an inverted pendulum system (denoted by \textsf{Inver}).

\subsection{Controller synthesis example: a remotely-controlled robot}
\begin{figure}
	\centering
	\includegraphics[width=0.80\textwidth]{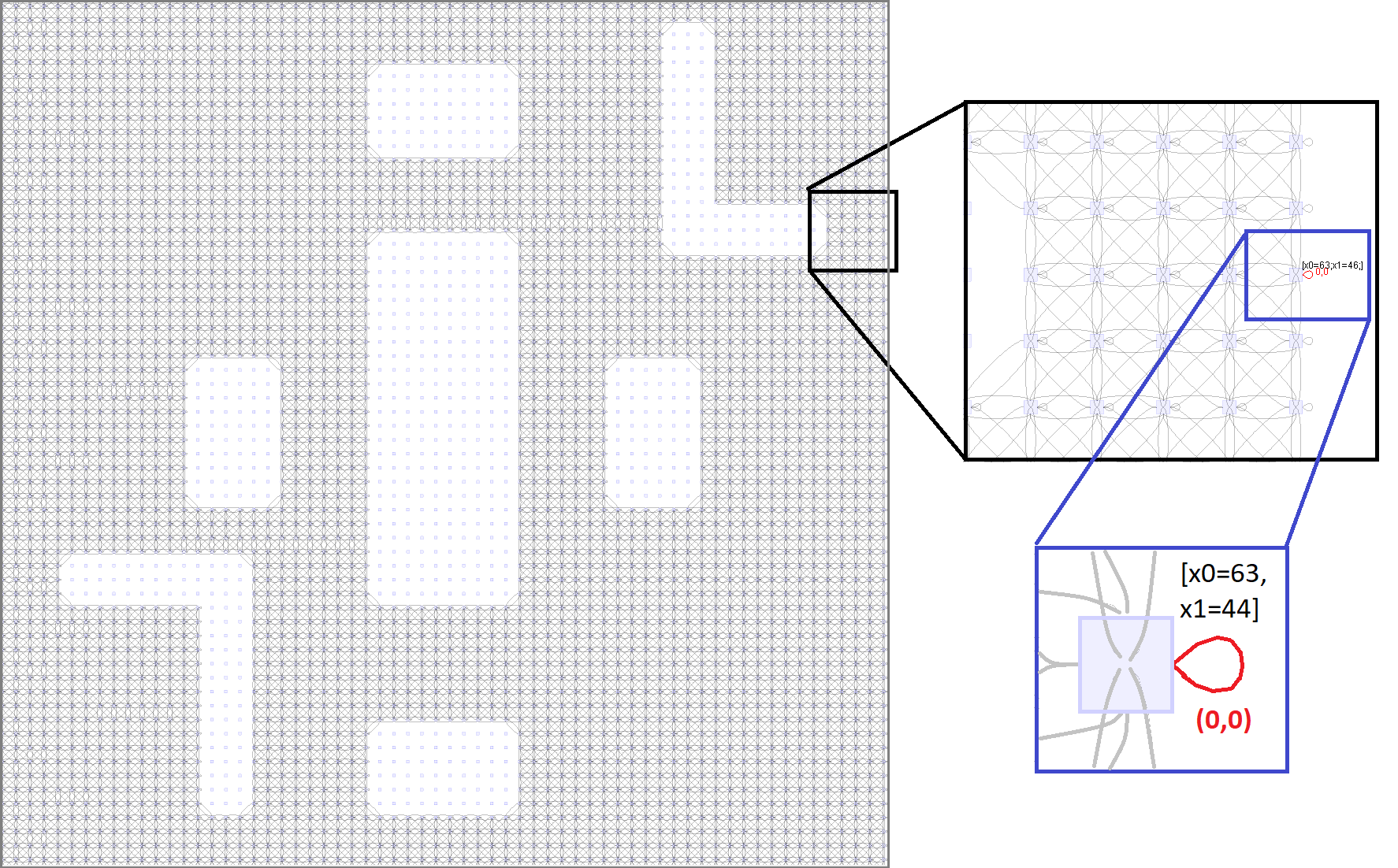}
	\caption{
		Visualization of the transition relation of the robot example. 
		In this example, obstacles in the state set are considered by removing any transitions to or from them.}
	\label{FIGROBOTTRANSREL}
\end{figure}

\begin{figure}
	\centering
	\subfloat[\MATLAB]{
		\centering
		\includegraphics[scale=0.263]{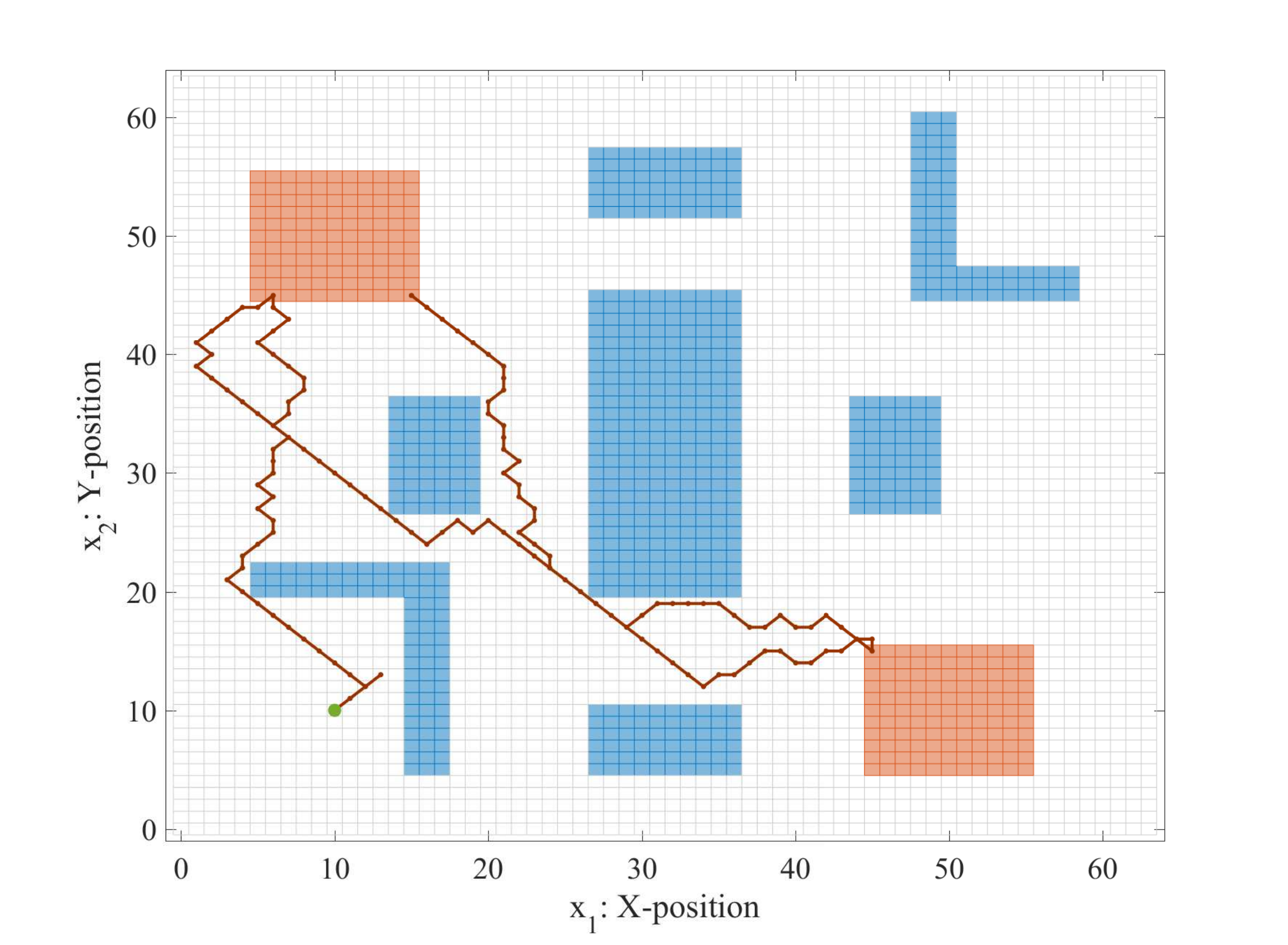}
		\label{FIGEXSIMMATLABROBOT}		
	}
	\subfloat[\OMNETPP]{
		\centering
		\includegraphics[scale=0.165]{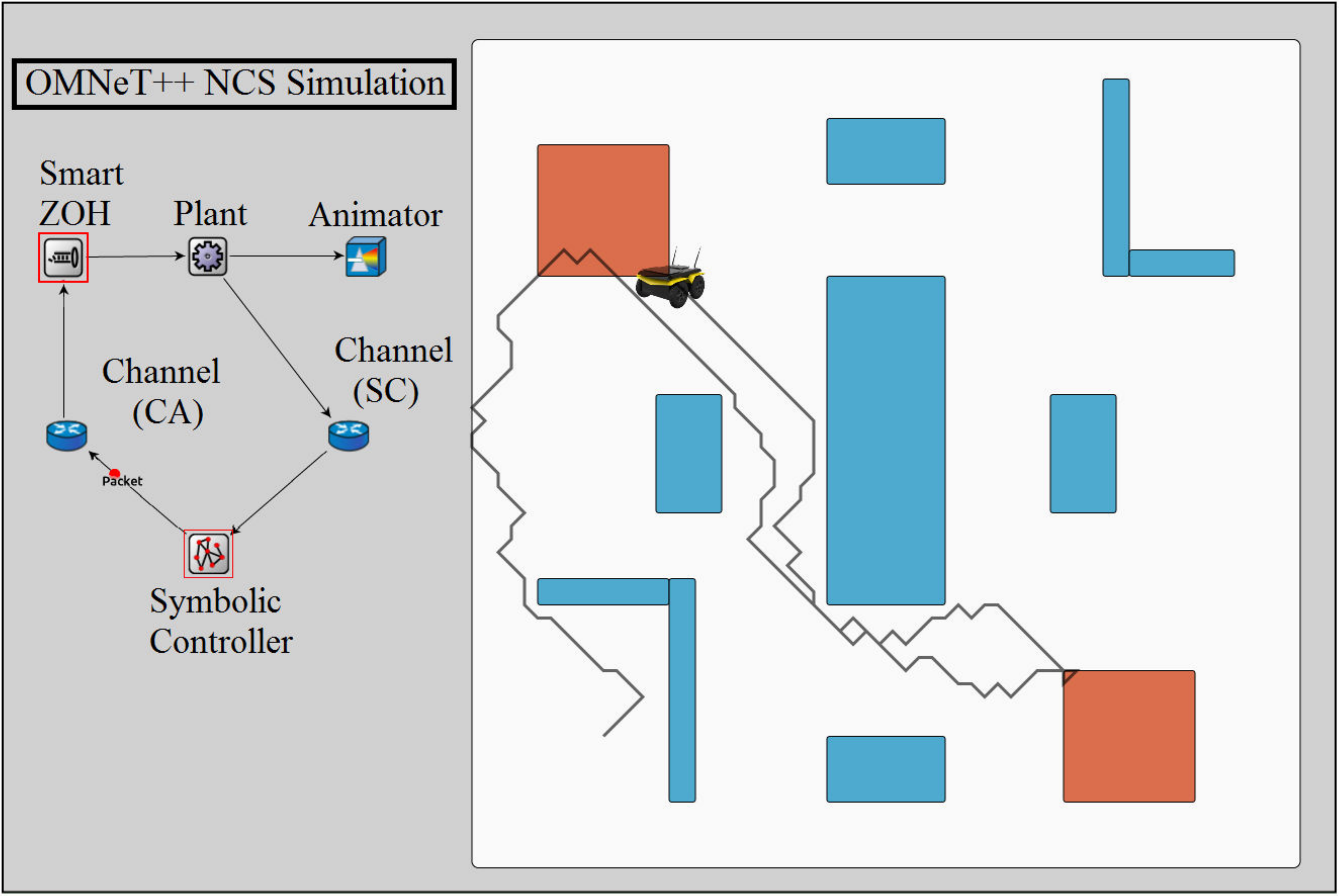}
		\label{FIGEXSIMOMNETROBOT}		
	}\\
	\caption{
		Closed-loop simulation of the robot in \MATLAB{} and in \OMNETPP. 
		Two target sets are represented by red boxes.
		Nine obstacle sets are represented by blue boxes.
	}
	\label{FIGEXSIM}
\end{figure}

For controller synthesis, we consider the \textsf{Robot} case reported in Table \ref{TBLMODELCONSDETAILS}.
We consider delay parameters of the network to be $(2, 2)$. 
The plant dynamic is described by the following differential equation:
\begin{equation}
\nonumber
\begin{bmatrix} 
\dot{\xi}_1\\ 
\dot{\xi}_2\\  				
\end{bmatrix}
=	
\begin{bmatrix} 
\upsilon_1\\ 
\upsilon_2\\ 				
\end{bmatrix},
\end{equation}
where $(\xi_1, \xi_2) \in X \subseteq \R^2$ denotes the position of the robot in the bounded 2D arena $X$, and $(\upsilon_1, \upsilon_2)$ represents a steering input.
We use the following state and input sets: set of states is $[0,64]\times[0,64]$, state quantization parameters are $(1,1)$, input set is $[-1,1]\times[-1,1]$, and input quantization parameters are $(1,1)$.
The control objective is described by the following LTL formula:
\begin{small}
	\vspace{-0.2cm}
	\begin{equation}
	\nonumber
	\psi = \Big( \underset{i=1}{\overset{9}{\bigwedge}} \square(\lnot \textsf{Obstacle}_i)\Big) \wedge \square\Diamond(\textsf{Target}1) \wedge \square\Diamond(\textsf{Target}2), 
	\end{equation}
\end{small}where the atomic propositions $\textsf{Target}1$, $\textsf{Target}2$, and $\textsf{Obstacle}_i$, $i\in\{1,\ldots,9\}$, are some hyper-intervals over $X$, as depicted in Figure \ref{FIGEXSIM}.

The library \SCOTS{} constructs the symbolic model of the plant in 0.23 seconds.
Then, the transition relation is exported using the tool {\tt bdd2fsm}.
Figure \ref{FIGROBOTTRANSREL} depicts the visualized transition relation using the generated FSM file and the tool StateVis.
The blank spaces in the state set represent obstacles. 

The NCS construction engine of \SENSE{} constructs the symbolic model of the NCS in 0.26 seconds.
The fixed-point operations engine in \SENSE{} synthesizes the controller in 8 seconds.
The resulting controller is a dynamic controller with two discrete states, each corresponds to one static controller, that enforces a reachability specification for one of the target sets.
Figure \ref{FIGEXSIMMATLABROBOT} (resp. \ref{FIGEXSIMOMNETROBOT}) shows the closed-loop simulation in \MATLAB{} (resp. \OMNETPP). 
We make use of the animation capabilities of \OMNETPP{} to visualize both packet transfers over the network as well as the movement of the robot in the arena.
The target sets are indicated with red boxes and the obstacles are indicated with blue boxes.
The initial state of the system is $(10,10)$.

\begin{figure}
	\centering
	\includegraphics[scale=0.45]{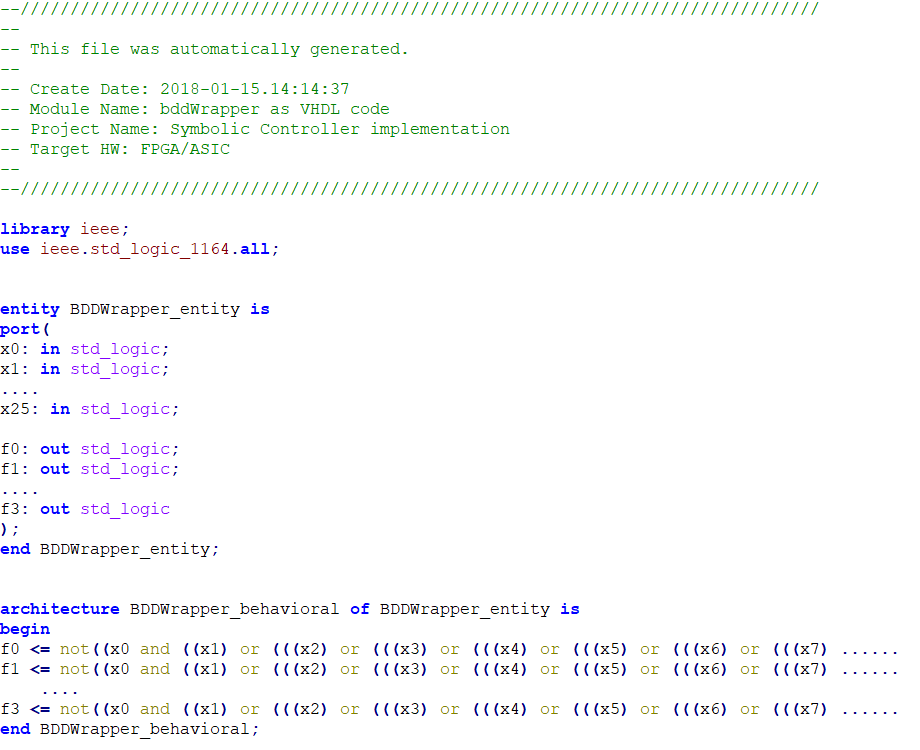}
	\caption{Code snippet from the generated VHDL code.}
	\label{FIGVHDLCODE}
\end{figure}

The tool {\tt bdd2implement} is used to generate VHDL and C/C++ codes that are ready for implementations.
Figure \ref{FIGVHDLCODE} shows a code-snippet of the generated VHDL code for one of the static reachability controllers of the \textsf{Robot} example.

\subsection{Controller synthesis example: collision-free deployment of two robots }
We use two instances of the same dynamic from the previous example to represent two different robots.
The dynamics are augmented to form a system of higher dimension (i.e. 4-dimensional state set and 4-dimensional input set).
The system is now described by the following differential equation:
\begin{equation}
\nonumber
\begin{bmatrix} 
\dot{\xi}_1\\ 
\dot{\xi}_2\\ 
\dot{\xi}_3\\ 
\dot{\xi}_4\\ 				
\end{bmatrix}
=	
\begin{bmatrix} 
\upsilon_1\\ 
\upsilon_2\\ 
\upsilon_3\\ 
\upsilon_4\\ 				
\end{bmatrix}.
\end{equation}

We use the following state and input sets: set of states is $[0,15]\times[0,15]\times[0,15]\times[0,15]$, state quantization parameters are $(1,1,1,1)$, input set is $[-1,1]\times[-1,1]\times[-1,1]\times[-1,1]$, and input quantization parameters are $(1,1,1,1)$.
The objective of this example is to synthesize a controller that enforces the following two specifications simultaneously:
\begin{enumerate}
	\item For the first robot:
	\begin{small}
		\begin{equation}
		\nonumber
		\psi_1 = 
		\Big( \underset{i=1}{\overset{5}{\bigwedge}} \square(\lnot \textsf{Obstacle}^1_i)\Big) 
		\wedge 
		\Big( \underset{i=0,j=0}{\overset{15,15}{\bigwedge}} \square(\lnot \textsf{Crash}_{i,j})\Big) 
		\wedge
		\square\Diamond(\textsf{Target}1) 
		\wedge 
		\square\Diamond(\textsf{Target}2),
		\end{equation}\end{small}
	
	\item For the second robot:
	\begin{small}
		\begin{equation}
		\nonumber
		\psi_2 = \Big( \underset{i=1}{\overset{5}{\bigwedge}} \square(\lnot \textsf{Obstacle}^2_i)\Big) 
		\wedge
		\Big( \underset{i=0,j=0}{\overset{15,15}{\bigwedge}} \square(\lnot \textsf{Crash}_{i,j})\Big) 
		\wedge
		\square\Diamond(\textsf{Target}3) 
		\wedge 
		\square\Diamond(\textsf{Target}4), 
		\end{equation}\end{small}
	
\end{enumerate}
where {\textsf Target$_1$}, {\textsf Target$_2$}, {\textsf Target$_3$} and {\textsf Target$_4$} are atomic propositions over the state set and defined by the hyper-rectangles:
$[2,3]\times[2,3]\times[0,15]\times[0,15]$, 
$[12,13]\times[12,13]\times[0,15]\times[0,15]$, 
$[0,15]\times[0,15]\times[2,3]\times[12,13]$,  and 
$[0,15]\times[0,15]\times[12,13]\times[2,3]$,  respectively.
$\textsf{Crash}_{i,j}$, $\forall i,j \in \{0,1, \dots, 15\}$, represent the points in the state set where the two robots collide and are defined as follows:
\begin{equation}
\nonumber
\textsf{Crash}_{i,j} = [i \; j \; i \; j]^T.
\end{equation}
$\textsf{Obstacle}^1_i$, $\forall i \in \{0,1, \ldots, 5\}$, are obstacles and are defined by the hyper-rectangle:
$[6,9]\times[6,9]\times[0,15]\times[0,15]$, 
$[11,14]\times[7,8]\times[0,15]\times[0,15]$, 
$[1,4]\times[7,8]\times[0,15]\times[0,15]$,
$[7,8]\times[1,4]\times[0,15]\times[0,15]$, and 
$[7,8]\times[11,14]\times[0,15]\times[0,15]$,  respectively.
$\textsf{Obstacle}^2_i$, $\forall i \in \{0,1, \ldots, 5\}$, are obstacles and are defined by the hyper-rectangle:
$[0,15]\times[0,15]\times[6,9]\times[6,9]$, 
$[0,15]\times[0,15]\times[11,14]\times[7,8]$, 
$[0,15]\times[0,15]\times[1,4]\times[7,8]$,
$[0,15]\times[0,15]\times[7,8]\times[1,4]$,
$[0,15]\times[0,15]\times[7,8]\times[11,14]$,  respectively.

\begin{figure}
	\centering
	\centering
	\includegraphics[scale=0.22]{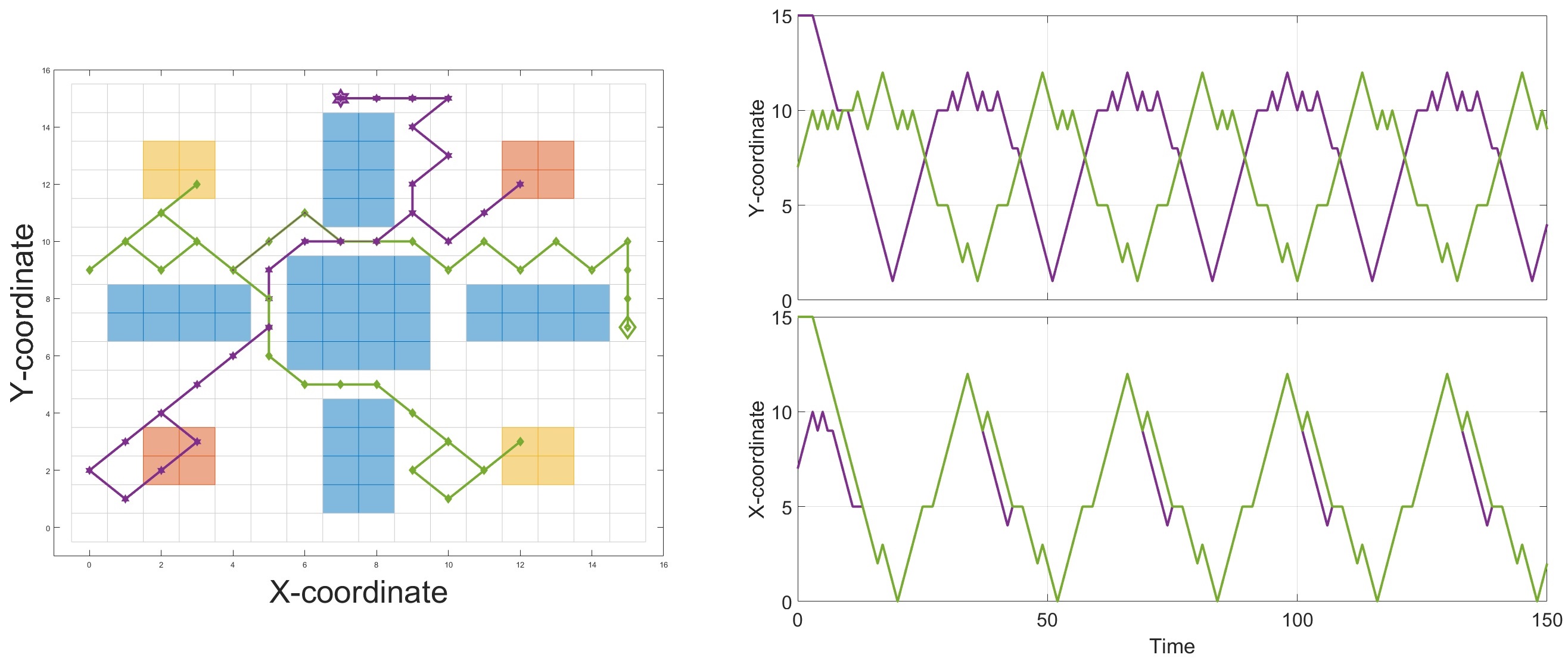}		
	\caption{Closed loop simulation of the two robots example in \MATLAB.}
	\label{FIGEXTWOBOTSMATLAB}
\end{figure}

The controller is synthesized and simulated as presented in the previous example.
The symbolic model of the plant is constructed in 56 seconds.
For delay parameters of $(2,2)$, \SENSE{} constructs the symbolic model of the NCS in 29 seconds.
Remark that the construction of the symbolic model of the plant in \SCOTS{} is affected by the increase in the dimension more than in \SENSE.
This is due to the nature of \SCOTS{} which operates element-by-element on the state set to construct an over-approximation of the reachable sets while constructing the finite abstraction of the plant.
On the other hand, \SENSE{} implements the $\mathcal{L}$-operator as operations on the BDD representing the symbolic model of the plant which is computationally efficient.
The tool \SENSE{} synthesizes the controller in 75883 seconds.
Figure \ref{FIGEXTWOBOTSMATLAB} shows the closed-loop simulation in \MATLAB{}. 
The initial state of the system is $(7,15,15,7)$.

\bibliographystyle{eptcs}
\bibliography{refdb}
\end{document}